\documentclass[12pt]{iopart}
\usepackage[english]{babel}
\usepackage{setstack}
\usepackage{graphicx}
\usepackage{epsfig}

\begin{document}

\title{Liquid crystals: a new topic in physics for undergraduates}
\author{Jerneja Pavlin$^1$, Nata\v{s}a Vaupoti\v{c}$^{2,3}$, Mojca \v Cepi\v
c$^{1,3}$}
\address{$^1$ Faculty of Education, University of Ljubljana, Slovenia} 
\address{$^2$ Faculty of Natural Sciences and Mathematics, University of
Maribor, Slovenia} 
\address{$^3$ Jo\v zef Stefan Institute, Ljubljana,
Slovenia} \ead{jerneja.pavlin@pef.uni-lj.si}

\begin{abstract}
The paper presents a teaching module about liquid crystals. Since liquid
crystals are linked to everyday student experiences and are also a topic of
a current scientific research, they are an excellent candidate of a modern
topic to be introduced into education. We show that liquid crystals can
provide a \textit{file rouge} through several fields of physics such as
thermodynamics, optics and electromagnetism. We discuss what students should
learn about liquid crystals and what physical concepts they should know
before considering them. In the presentation of the teaching module that
consists of a lecture and experimental work in a chemistry and physics lab,
we focus on experiments on phase transitions, polarization of light, double
refraction and colours. A pilot evaluation of the module was performed among
pre-service primary school teachers who have no special preference for
natural sciences. The evaluation shows that the module is very efficient in
transferring knowledge. A prior study showed that the informally obtained
pre-knowledge on liquid crystals of the first year students on several
different study fields is negligible. Since the social science students are
the ones that are the least interested in natural sciences it can be
expected that students in any study programme will on average achieve at
least as good conceptual understanding of phenomena related to liquid
crystals as the group involved in the pilot study.
\end{abstract}

\maketitle

\section{Introduction}

For physicists physics is a permanent inspiration for new discoveries.
However, non-physicists often consider physics as a boring and old
discipline, detached from everyday life. Public often fails to realize the
consequences of research in everyday applications, so it often considers the
academic research as a waste of financial resources. But research is tightly
connected to the development even if it is not strongly focused toward
applications. This can be best illustrated by the well known statement that
\textquotedblleft the light bulb was not discovered by optimizing a candle\textquotedblright \cite{Ghidaglia,
Sarkozy}. The apparent non-relevance of physics for the everyday life is
often caused by the choice of topics taught during the lectures, which are
usually old from the point of young students, since even the most recent
topics - fundamentals of modern physics - are more than a hundred years old 
\cite{shabajee}. In addition, traditional teaching very often considers
idealized examples and, worst of all, present experiments as a \textquotedblleft proof\textquotedblright for
theoretical explanations.

The physics education research has pointed out several of these problems and
the physics education in general has advanced tremendously in the last
twenty years \cite{Osborne_2003}. But topics that introduce a part of the
frontier research into the classroom, showing the students that the physics
is not a dead subject yet, are still extremely rare.

In this paper we present a topic, liquid crystals, which is one of rare
examples, where such a transfer is possible. The community occupied by the
research on liquid crystals counts several thousands of researchers. We all
experience the consequences of research on liquid crystals every day; every
mobile phone, every portable computer and almost every television screen is
based on the technology using liquid crystals.

The physics of liquid crystals is not very simple but there are several
concepts that can be understood by non-physics students as well, especially
if the teaching approach is based on gaining practical experiences with
liquid crystals. In addition, for advanced levels of physics students,
liquid crystals may serve as a clear illustration of several concepts
especially in thermodynamics and optics.

A serious interest of researchers for an introduction of liquid crystals
into various levels of education was first demonstrated at The International
Liquid Crystal Conference (ILCC) in Krakow, Poland, in 2010. ILCC is a
biennial event gathering more than 800 researchers studying liquid crystals
from a variety of aspects. In Krakow, one of four sections running in
parallel was called \textit{Liquid Crystals in Education}. The audience
unexpectedly filled the auditory to the last corner and after lectures
lengthy discussions developed \cite{cepic_2010}. A similar story repeated at
the education section at the European Conference on Liquid Crystals in
Maribor, Slovenia, in 2011, and at ILCC in Mainz, Germany, in 2012.

At present, some of the physics of liquid crystals is usually briefly
mentioned at various courses at the university level, but there is no
systematic consideration from the education perspective about the importance
of various concepts and teaching methods. To our best knowledge, there exist
no example of a model teaching unit. In this contribution we report on a
teaching module on liquid crystals, which is appropriate for the
undergraduate level for non-physicists. The module can be extended to the
lab work at more advanced levels. Most of the module can also be used in
courses related to thermodynamics and optics as demonstration experiments or
lab work accompanied by more rigorous measurements and calculations, which
are not considered in detail in this contribution.

The paper is organized as follows: in section 2 we consider the
prerequisites for the introduction of new modern topic into education.
Before designing a module we had to consider several points, not necessary
in the same order as quoted here: What outcomes do we expect of the teaching
module? Which are the concepts that students should understand and be able
to apply after the module? Where in the curriculum should the topic be
placed, or equivalently, what is the knowledge students need to be able to
construct new knowledge about liquid crystals? Which teaching methods are
most appropriate for the teaching module? And finally, do we have all the
materials like experiments, pictures, equipment and facilities to support
the teaching module? In section 3 we report the pilot evaluation study of
the teaching module, which was performed in 2011. In section 4 we conclude
and discuss several opportunities that the new teaching module offers to the
physics education research in addition to the new knowledge itself.

\section{Teaching module}

When we consider a new topic which is a part of contemporary research with
applications met every day, and we want to adapt it for teaching purposes,
the literature search is not much of a help. A thorough literature search
did not show any theoretical frameworks on this topic. One can find
theoretical frameworks for various approaches to teaching and
discussions about students motivation and understanding of various concepts.
We have found few examples of introduction of new topics like an
introduction of semiconductors into the secondary school or introduction of
more advanced concepts with respect to friction only \cite{Garciacarmona,
Besson_2007}. There are also examples of introduction of concepts of quantum
mechanics into high school \cite{Escalada_1997, Cuppari_1997, Muller_2002,
olsen_2002}. All authors reported similar problems with respect to the
existing theories and results in physics and science education research;
they had to build the units mostly from the personal knowledge, experience
and considerations. On the other hand, several approaches for analytical
derivation of already treated concepts, several suggestions for
demonstrations and lab experiments for teaching purposes are published in
every issue of the American and European Journal of Physics. This simply
means that the physics community is highly interested in the improvement of
the teaching itself, but the motivation of the researchers, being also
lecturers, lies more in the area of developing new experiments than in
thorough studies of their impact. Therefore, a lot of material for teaching
purposes for any topic, old, modern or new is available, but one further
step is usually needed towards the coherent teaching module.

With the above mentioned problems in mind, we begin by brief discussion on
what liquid crystals are and then give a short overview of the existing
literature regarding the introduction of liquid crystals into teaching. Then
we focus on the teaching module: we define our goals, consider the
pre-knowledge on which the module should be built and finally describe
details of the module.

\subsection{What are liquid crystals?}

Liquid crystals are materials which have at least one additional phase
between the liquid and the solid phase. This phase is called the liquid
crystalline phase and it has properties of both the liquid and the
crystalline phase: a) it flows like a liquid, or more fundamental, there is
no long range order in at least one of directions, and b) it is anisotropic,
which is a property of crystals, or, again, more fundamental, there exists a
long range order in at least one of directions. The name liquid crystal is
the name for the material, which exhibits at least one liquid crystalline
phase \cite{degennes_1993}.

\begin{figure}[h]
\centering \includegraphics[scale=0.6]{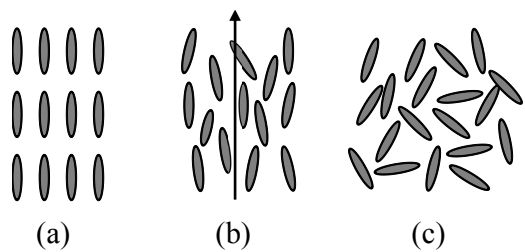} 
\caption{Molecules of liquid crystal in different phases; (a) crystalline,
(b) liquid crystalline (nematic) and (c) isotropic liquid.}
\label{fig:fig_1}
\end{figure}

Liquid crystalline phases differ by the way of long range ordering. In this
contribution we will discuss only the simplest type of ordering that is
typical for the nematic phase. Its properties are applied in a liquid
crystalline screen. The molecular order in the crystalline phase is shown
schematically in figure \ref{fig:fig_1} (a), in the nematic liquid
crystalline phase in figure \ref{fig:fig_1} (b) and in the isotropic liquid
phase in figure \ref{fig:fig_1} (c). One can see that, in the liquid
crystalline phase, there exists some orientational order of long molecular
axes. Such a material is anisotropic, the physical properties along the
average long molecular axis obviously being different from the properties in
the direction perpendicular to it. There are several liquid crystal phases
made of molecules having rather extravagant shapes. However, we shall limit
our discussion to the simplest case of nematic liquid crystal made of
rod-like molecules without any loss of generality of the phenomena studied
within the teaching module.

\begin{figure}[h]
\centering \includegraphics[scale=0.6]{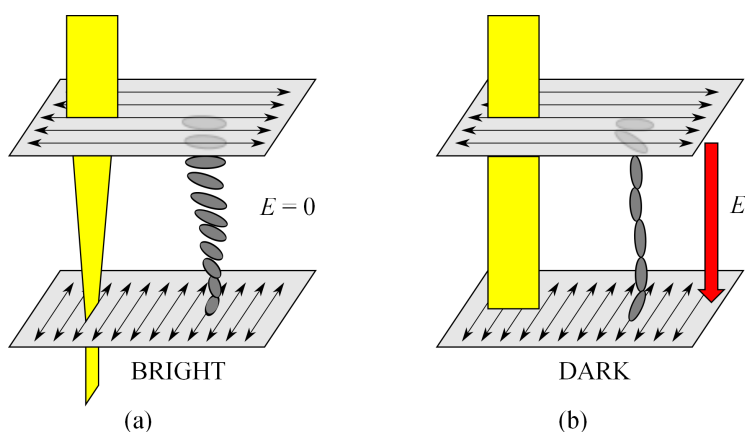} 
\caption{The arrangement of molecules in cells used for displays. (a) The
\textquotedblleft bright state\textquotedblright: without an applied
electric field ($\mathit{E}$ = 0) the cell transmits light; (b) the
\textquotedblleft dark state\textquotedblright: when voltage is applied to
the two glass plates molecules rearrange and the cell does not transmit
light.}
\label{fig:fig_2}
\end{figure}

When liquid crystal molecules are close to surfaces, surfaces in general
tend to prefer some orientation of long molecular axes. Using a special
surface treatment one can achieve a well-defined orientation of long
molecular axes at the surface, for example, in the direction parallel to the
surface or perpendicular to it. In liquid crystal cells, which are used in
liquid crystal displays (LCDs), surfaces are usually such that they anchor the
molecules by their long molecular axes in the direction parallel to the
surface; however orientations of molecules at the top and bottom surface are
perpendicular. Molecules between the surfaces tend to arrange with their
long axis being parallel, however due to the surface anchoring their
orientation rotates through the cell (figure \ref{fig:fig_2} (a)). Because
in the anisotropic materials the speed of light depends on the direction of
light propagation and on the direction of light polarization, liquid
crystals organized in such a special way rotate direction of light
polarization. If such a cell is put between two crossed polarizers whose
transmission directions coincide with the direction of surface anchoring,
the cell transmits light \cite{degennes_1993, collings_1997}.

Liquid crystals are also extremely useful in discussing different competing
effects that determine the molecular arrangement. Application of an external
magnetic or electric field changes the structure in the liquid crystal cell
described above, because the electric or magnetic torque tends to arrange
molecules in the direction parallel or perpendicular to the external field,
depending on the molecular properties. Rotation of molecules in the external
field changes the optical transmission properties of the cell. This is a
basis of how LCDs work: with the field on (figure \ref%
{fig:fig_2} (b)), light is not transmitted through a cell and the cell is
seen to be dark; with the field off (figure \ref{fig:fig_2} (a)), light is
transmitted through the cell and the cell is seen to be bright.

Because of their unique physical properties several authors have already
considered the introduction of liquid crystals into the undergraduate
university studies. A mechanical model of a three dimensional presentation
of liquid crystals phases is presented in \cite{repnik_2011}. The historical
development of the liquid crystals research and their application is given
in \cite{ondris_1992}. In \cite{crawford_1994} the same authors point out
that liquid crystals are an excellent material to connect some elementary
physics with technology and other scientific disciplines.

The procedures to synthesize cholesteric liquid crystals (nematic liquid
crystals in which the average orientation of long molecular axes spirals in
space) in a school lab at the undergraduate university level are given in 
\cite{patch_1985, hecke_2005} together with the methods to test the
elementary physical properties of liquid crystals. In the Appendix to \cite%
{hecke_2005} an experiment to determine refractive indices of a liquid
crystal is discussed \cite{shenoy}. The anisotropic absorption of polarized
light in liquid crystals is presented.

Liquid crystals that are appropriate for the education purposes are
thermotropic; their properties change with temperature. The colour of
cholesteric liquid crystals changes if temperature increases or decreases,
so they can be used as thermometers \cite{Lisensky_2005}. They are also
sensitive to pressure. Reference \cite{Katzl} contains worksheets for an
experiment in which students discover the sensitivity of cholesteric liquid
crystals mixtures on pressure and temperature. In \cite{Olah_1977} a simple
experimental setup is presented by which students can detect and record the
light spectra, study and test the concept of Bragg reflection, and measure
the anisotropy of a refractive index in a cholesteric liquid crystal.

A series of simple experiments that can be shown during lectures and can
bring the science of liquid crystals closer to students is described in \cite%
{Ciferno_1995}. The experiments are used to introduce the concepts of
optics, such as light propagation, polarization of light, scattering of
light and optical anisotropy. Liquid crystal can also be used to describe
light transmission through polarizers \cite{Evans_2008}. When an external
field is applied to a cell, a threshold value is required to rotate
molecules in the direction preferred by the field. This effect is called the
Freedericksz transition and an experiment for the advanced physics lab at
the undergraduate level is presented in \cite{Moses_1998}.

The procedure to prepare a surface-oriented liquid crystal cell is given 
\cite{Liberko_2000} where the procedure to synthesize a nematic liquid
crystal 4-methoxybenzylidene-4'-n-butylaniline (MBBA) is provided as well.
Several experiments that illustrate optical properties of liquid crystals
are shown. One learns how to design a cell in which molecules are uniformly
oriented and what is observed, if this cell is studied under the polarizing
microscope. A detailed description of phase transitions between the liquid,
liquid crystal and crystal phases is given. Exercises are interesting for
undergraduate students because they synthesize the substance which they use
for other experiments, e.g. measurements of the refractive indices.

There are several advanced articles which give advice on the inclusion of
liquid crystals into the study process at the university, both undergraduate
or graduate, level. An experiment for the advanced undergraduate laboratory
on magnetic birefringence in liquid crystals is presented in \cite%
{Moses_2000}, measurements of order and biaxiality are addressed in \cite%
{low_2002}. In \cite{Repnik_2003} defects in nematic liquid crystals are
studied by using Physics Applets \cite{kaucic_2004}. Liquid crystals are not
used only in displays but also in switches. In \cite{Boruah} a liquid
crystal spatial light modulator is built and used for a dynamic manipulation
of a laser beam.

\subsection{What should students learn about liquid crystals?}

When discussing an introduction of a new topic into education, the team
should be aware of the goals as well as of limitations. There are no general
criteria established about the concepts that student should learn and
comprehend, when a new topic is introduced. Therefore, for liquid crystals,
we had to rely on our understanding of the topic; we had to neglect our
personal bias toward the matter as two of the authors are also active
researchers in the theoretical modeling of liquid crystals. Being an expert
in the research of liquid crystals may also over or under estimate the
concepts that are important for students. We hope that we managed to avoid
these \textquotedblleft personal\textquotedblright traps.

The aim of the teaching module is to explain how LCDs work and what is the role of liquid crystals in its operation. We
also believe that some basic understanding of LCDs is a part of a general
public knowledge, because it is an important example of the link between the
academic research and the applications, which follow from it. If non-physics
students meet such an example at least once during their studies, they might
not consider the academic research that has no immediate specific
application as obsolete.

According to our opinion\ students should obtain and understand the
following specific concepts:

\begin{itemize}
\item they should be able to recognize and identify the object of interest -
the pixel - on an enlarged screen;

\item they should be aware of the fact that liquid crystals are a special
phase of matter having very special properties;

\item they should become familiar with the following concepts: anisotropy,
double refraction and birefringence;

\item they should be aware of the fact that liquid crystals must be ordered
if we want to exploit their special properties;

\item they should know that liquid crystals are easily manipulated by
external stimuli like an electric field; and finally

\item they should link the concepts mentioned above in a consistent picture
of pixel operation.
\end{itemize}

Before constructing the teaching module we have thus set the goals stated
above. The next steps are: a) to consider the required knowledge before the
students start the module, b) to position the topic in the curriculum,
otherwise teachers will probably not adopt it and c) to choose the methods,
which will be the most successful for constructing the new knowledge.

\subsection{Which basic physics concepts should students know before
considering liquid crystals?}

As liquid crystals are materials which form a special liquid crystalline
phase, students should be familiar with a concept of phases. They should be
aware of a phase transition that appears at an exact temperature. Students
very often believe that mixture of ice and water can have any temperature as
they often drink mixtures of ice and water, which are not in a
thermodynamically stable state yet.

The next important concept is the speed of light in a transparent medium,
its relation to the index of refraction and Snell's law. If students are
familiar with the concept of polarized light and methods of polarizing the
light, it is an advantage. However, polarization of light can also be nicely
taught when teaching the properties of liquid crystals.

Students should also know, at least conceptually, that materials
electrically polarize in an external electric field and that the material
polarization (not to be confused with polarization of light!) is a
consequence of structural changes of a material in the electric field.

Considerations of the required preliminary students knowledge also give some
hints about the placement of the topic on liquid crystals in the curriculum:

\begin{itemize}
\item When teaching thermodynamics or, more specifically, phases and phase
transitions, additional phase can be visually shown by liquid crystals,
since the appearance of liquid crystals in their liquid crystalline phase is
significantly different from their solid or liquid state. This is not the
case for other materials, for example, a ferromagnetic material or a
superconductor looks exactly the same when the phase transition to the
ferromagnetic or superconductive phase appears at the transition
temperature. The phase transition has to be deduced from other properties.

\item When Snell's law is introduced and prisms are discussed, a rainbow is
often added as an interesting phenomenon that is observed because the speed
of light depends on its wavelength. Similar phenomenon, i.e. a dependence of
the speed of light on light polarization can be discussed by a phenomenon of
a double refraction.

\item One picture element (pixel) is formed by confining LC between two
conducting plates. One pixel is thus a capacitor with a dielectric material
(liquid crystal) between the plates. When voltage is applied to the cell
surfaces (capacitor plates), the material between the plates polarizes.
Electric polarization leads to changes in structural properties of the
materials, in this case to the reorientation of molecules, which affects the
transmission of light through the cell. Thus a LC pixel can be used to
consider electric polarization of other materials that can structurally
change due to the reorientation of molecules.
\end{itemize}

From the above we clearly see that liquid crystals can provide a
motivational \textit{file rouge} through several topics. On the other hand,
by showing several phenomena related to liquid crystals, teachers can
motivate students to remain interested in various topics in physics and link
them together in the explanation of how one pixel in a liquid crystal
display works. Teachers can choose to teach about liquid crystals as a
separate topic aiming to establish a link between a current fundamental
research topic and consequences of the research applied and used every day
by everybody.

The last question that remains to be answered is a choice of methods for the
teaching intervention. Liquid crystals are a new topic for students. They
are mostly only slightly familiar with the name and have practically no
associations connected to the name except a loose connection to displays, as
will be shown later. Therefore the topic should be introduced from the very
beginning. Due to several concepts that must be introduced and the structure
of understanding that students have to build without any pre-knowledge,
traditional lecturing seems the most natural choice. However, from the
literature and our experiences we are aware that the transfer of knowledge
to relatively passive students is not as successful as one would wish for.
Therefore we decided to use a combination of a traditional lecture
accompanied by several demonstration experiments, where most of the
fundamental concepts and properties are introduced, a chemistry lab, where
students synthesize a liquid crystal, and a physics lab, where they use
their own product from the chemistry lab to study its various physical
properties by using an active learning approach. The lab work allows
students to construct and to comprehend several new ideas that are all
linked together in the application, a liquid crystalline display.

\subsection{The teaching module}

The teaching module gives the basic knowledge about liquid crystals, which
we assessed as necessary for the understanding of liquid crystals and liquid
crystal display technology for a general citizen having at least a slight
interest in science and technology. The teaching module has three parts:
lecture (1st week), lab work at chemistry (2nd week) and lab work at physics
(3rd week). The estimated time for each part is 90 minutes. Within the
teaching module we wanted the students to assimilate the following concepts:
a) a synthesis of a liquid crystals MBBA \cite{verbit_1972, Pavlin_2011},
b) an existence of an additional phase and phase transitions, c)
polarization of light and d) optical properties of liquid crystals related
to anisotropy \cite{Pavlin_2011}. Below we present the module, its aims, its
structure and a short description of activities in each part of the teaching
module.

\subsubsection{Lecture}

The lecture in duration of 90 minutes provides the fundamental information
about liquid crystals, about their properties and how they are used in
applications. The method used is a traditional lecture accompanied by
several demonstration experiments that are used for motivation, as a
starting point for discussion and as an illustration of phenomena discussed.

After the lecture students should be able to

\begin{itemize}
\item list some products based on the liquid crystal technology;

\item recognize the additional liquid crystalline phase and phase transition;

\item describe and illustrate the structure of liquid crystals on a
microscopic level;

\item recognize the properties of liquid crystals, which are important for
applications: birefringence, resulting from the orientational ordering of
molecules, and the effect of an electric field on molecular orientation;

\item describe how a LCDs work;

\item know that liquid crystals are also found in nature and that they are
present in living organisms.
\end{itemize}

In addition, a short part of the lecture introduces polarizers and their
properties, since most of the students have not heard of the concept of
polarization and polarizers during their previous education.

The lecture starts with a magnification of a LC screen as a motivation; it
is explained that at the end of the module students will be able to
understand how the display works. The lecture continues with a description
and a demonstration of the new, liquid crystalline, phase, the macroscopic
appearance of which is similar to an opaque liquid. All three phases
(crystalline, liquid crystalline and isotropic liquid) are shown while
heating the sample. The microscopic structures of all three phases are
presented by cartoons and the orientational order is introduced. The
molecular shape which allows for the orientational ordering is discussed.
The concept of light propagation in an anisotropic material is introduced
and double refraction is shown by using a wedge liquid crystalline cell.
Colours of an anisotropic material (scotch tape) between crossed polarizers
are demonstrated and explained. When polarized light propagates through an
anisotropic material, the polarization state of light, in general, changes
from the linearly polarized to elliptically polarized. The state of
elliptical polarization is defined by the wavelength of light, the
birefringence of the anisotropic material (i.e. the difference between the
refractive indices) and the thickness of the material. The understanding of
how polarized light propagates through a birefringent (optically
anisotropic) material is crucial for understanding the pixel operation.

In the LCDs the electric properties of molecules are very
important so they have to be introduced in the lecture. The effect of the
electric field on the molecular orientation is discussed as well. Molecules
are described as induced electric dipoles that are rotated by the external
electric field. Because the anisotropic properties depend on the structure
of the liquid crystal in the cell, the transmission depends on the applied
electric field. This leads to the structure of a pixel and to how liquid
crystal displays work.

At the end of the lecture some interesting facts are mentioned, such as
liquid crystals being a part of spider threads and cell membranes in living
organisms.

\subsubsection{ Lab work in chemistry: synthesis of liquid crystal MBBA}

The aims of the lab work in chemistry are the following:

\begin{itemize}
\item Students are able to synthesize the liquid crystal MBBA.

\item Students realize that the product of the synthesis is useful for the
experiments showing the basic properties of liquid crystals.
\end{itemize}

Students synthesize the liquid crystal MBBA in a school lab from
4'-n-butylaniline and of 4-methoxybenzaldehyde \cite{Pavlin_2011}. Due to the
safety reasons, the synthesis has to be carried out in the fume hood. This
part of the teaching module can be left out if the laboratory is not
available and the lab work in physics could extend to two meetings of the
duration of 90 minutes, which allows for more detailed studies of phenomena.

\subsubsection{ Lab work in physics}

Four experiments that are carried out during the lab work in physics provide
students with personal experiences and allow them to investigate the most
important liquid crystalline properties.

\textbf{Experiment 1: An additional phase and phase transition }

Aims:

\begin{itemize}
\item Students know that the liquid crystalline state is one of the states
of mater.

\item Students are able to describe the difference between the melting
temperature and the clearing temperature.

\item Students are able to measure these two temperatures and use them as a
measure of the success of the synthesis. If both temperatures are close to
the temperatures given in the published data, the synthesis was successful.
\end{itemize}

\begin{figure}[h]
\centering \includegraphics[scale=0.4]{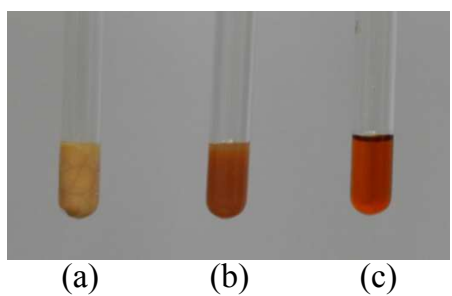} 
\caption{The phases of the liquid crystal MBBA: (a) solid, (b) opaque liquid
crystalline and (c) clear isotropic liquid.}
\label{fig:fig_3}
\end{figure}

Students use a water bath to heat the test tube with a frozen liquid crystal
MBBA. They measure the temperature of water assuming that the small sample
of liquid crystal has the same temperature as the bath. They observe how the
appearance of the substance changes (figure \ref{fig:fig_3}) while heating
the water bath. They measure the temperature at which the sample begins to
melt. This temperature is called the melting temperature and it is the
temperature of the phase transition from the crystalline to the liquid
crystalline phase. Students heat the water bath further and measure the
temperature at which the milky appearance of the sample starts to disappear.
This temperature is called the clearing temperature and it is the
temperature of the phase transition from the liquid crystalline phase to the
isotropic liquid.

\textbf{Experiment 2: Polarization}

Aims:

\begin{itemize}
\item Students know what polarizers are and how they affect the unpolarized
light.

\item Students know how light propagates through the system of two
polarizers.

\item Students are able to test, if the light is polarized and in which
direction it is polarized by using the polarizer with a known polarizing
direction.

\item Students are able to test if the substance is optically anisotropic by
using two polarizers.
\end{itemize}

Students use two polarizers and investigate the conditions under which light
propagates through two polarizers or is absorbed by them. They compare the
transmitted light intensity as a function of the angle between the
polarizing directions of polarizers. By this part of the activity they learn
how to use a polarizer as an analyzer. They also verify that the reflected
light is partially polarized.

\begin{figure}[h]
\centering \includegraphics[scale=0.6]{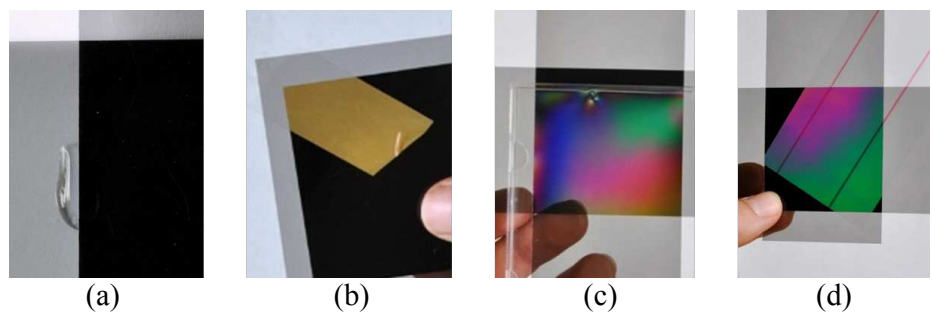} 
\caption{Various materials between crossed polarizers: (a) an isotropic drop
of water, (b) a piece of a scotch tape, (c) a transparent CD box and (d) a
cellophane.}
\label{fig:fig_4}
\end{figure}

Students observe various transparent materials placed between crossed
polarizers and find that light cannot be transmitted when (isotropic)
materials like water or glass are placed between the crossed polarizers.
When some other material like a scotch tape, cellophane or CD box is placed
between the crossed polarizers, light is transmitted. Colours are also often
observed (figure \ref{fig:fig_4}). Such materials are anisotropic. Students
can investigate how various properties of anisotropic materials (thickness,
type of material) influence the colour of the transmitted light. The
activity provides an experience that is later used for observations of
liquid crystals in a cell.

\textbf{Experiment 3: Double refraction}

Aims:

\begin{itemize}
\item Students know that birefringence is an important property of matter in
the liquid crystalline state.

\item Students are able to make a planar wedge cell and find an area where
liquid crystal is ordered enough that the laser beam splits into two
separate beams. The beams are observed as two light spots on a remote screen.

\item Students know how to check light polarization in a beam by a polarizer.
\end{itemize}

\begin{figure}[h]
\centering \includegraphics[scale=0.6]{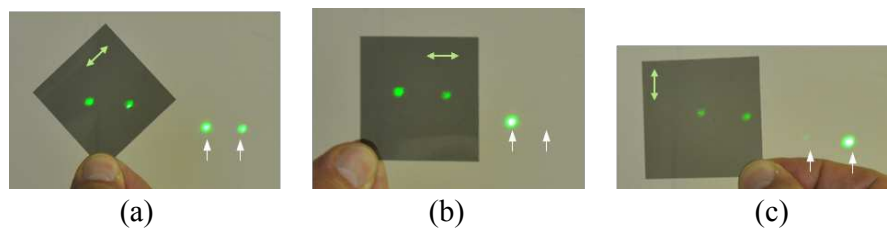} 
\caption{Finding the polarization direction of beams transmitted through the
LC wedge cell. The arrow on the polarizer marks the polarizing direction of
the polarizer. The polarizer transmits (a) both beams, (b) only the
extraordinary beam, (c) only the ordinary beam. Polarization of the ordinary
beam is perpendicular to the polarization of the extraordinary beam. Two
small arrows marking the light spots are used as a guide to the eye.}
\label{fig:fig_5}
\end{figure}

Students manufacture a wedge cell from a microscope slide, a cover glass, a
foil for food wrapping or a tape and the liquid crystal MBBA \cite%
{Pavlin_2011}. A special attention is given to the rubbing of the microscope
and cover glass, which enables anchoring of the liquid crystal molecules.
The rubbing also prevents the disorder of clusters of molecules with the
same orientation of long molecular axis; such clustering results in
scattering of light and opaqueness. Students direct the light on the wedge
cell (they use a laser pointer as the light source) and find the area of the
cell where the laser beam splits into two beams. By rotating the polarizer
between the cell and the screen they verify light polarization in the beams
(figure \ref{fig:fig_5}). Then students heat the wedge cell with the
hair-dryer and observe the collapse of the two bright spots into one at the
phase transition to the isotropic liquid.

\textbf{Experiment 4: Colours}

Aims:

\begin{itemize}
\item Students are able to fabricate a planar cell, i.e. a cell with
parallel glass surfaces.

\item Students know that liquid crystals are optically anisotropic and that
light is transmitted if a cell filled with liquid crystal in its liquid
crystalline phase is placed between two crossed polarizers.They know that
under such circumstances colours may also appear when the sample is
illuminated by white light.

\item Students know that the colours observed under perpendicular and under
parallel polarizers are complementary.

\item Students are able to mechanically order molecules in a planar cell.
\end{itemize}

\begin{figure}[h]
\centering \includegraphics[scale=0.6]{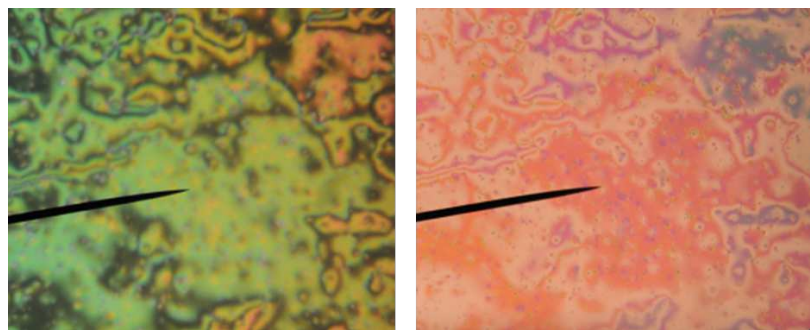} 
\caption{A planar cell with non-ordered molecules of MBBA under a polarizing
microscope. The polarizers are perpendicular (left) and parallel (right).}
\label{fig:fig_6}
\end{figure}

Students manufacture a planar cell filled with the liquid crystal MBBA from
a microscope slide, a cover glass and a foil for food wrapping or a tape 
\cite{Pavlin_2011}. They observe the planar cell under a polarizing
microscope (a school microscope with $M=40$ or $100$ and two crossed
polarizing foils). At this point students should remember the descriptive
definition of optically anisotropic materials and find out that liquid
crystals are their representatives. Then they rotate the polarizing foils
and observe how colours change (see figure \ref{fig:fig_6}). This experiment
also illustrates a concept of complementary colours \cite{babic_2009}.
Afterwards students heat the cell and observe colour changes. An experiment
where molecules are ordered is done next. Students make micro notches on a
microscope slide by rubbing it by a velvet soaked in alcohol. Molecules
orient with their long axes parallel to the surface and the rubbing
direction. A similar process is used in the fabrication of liquid crystal
displays. They observe the cell with the ordered liquid crystal under the
polarizing microscope.

Finally, the cell is heated by a hair-dryer and students observe the phase
transition that appears as a dark front moving through the sample ending in
a dark image when the liquid crystal between crossed polarizers is in the
isotropic phase. The lab work is concluded by a discussion of how one pixel
in the LCD works, relating the changes of the liquid
crystalline structure due to the electric field to the transmission rate of
the pixel. At this point the fact that colour filters are responsible for
colours of each part of the pixel is also emphasized.

\section{Pilot evaluation of the teaching module}

The aim of our study was development of a teaching module for non-physics
students, which could also be implemented for the high school students. Our
goal was to give future primary school teachers basic knowledge about liquid
crystals, so that they will be able to answer potential questions of younger
students when they will be teachers themselves. Therefore, the teaching
module described in the previous section was preliminary tested by a group
of 90 first year students enrolled in a four-year University program for
Primary school teachers at the Faculty of Education (University of
Ljubljana, Slovenia) in the school year 2010/11.

In this section we present the evaluation of the module as regards the
efficiency of teaching intervention: which concepts do students assimilate
and comprehend and to what extent?

\subsection{Methods}

\subsubsection{Participants}

First-year pre-service teachers (future primary school teachers) were chosen
for testing the teaching module. They were chosen, because their
pre-knowledge on liquid crystals is just as negligible as the pre-knowledge
of students from other faculties and study programs (see section 4). In
addition, the pre-service teachers do not have any special interest in
natural sciences, but they have to be as scientifically literate as everyone
else who has finished high school. And most important, the pre-service
primary school teachers form the only homogeneous group that has Physics
included in the study program and is, at least approximately, large enough to allow for a
quantitative study. In the group of $90$, $6$ were male and $84$ female
students. They were on average $20.1$ years old ($SD=1.6$ years). On
average, they achieved $19.7$ points out of $34$ ($SD=3.6$) on the final
exam at the end of the secondary school. The average achievement on the
final exam in Slovenia was $19.5$ points out of $34$ and a total of $8842$
candidates attended the final exams in spring 2010. The studied group
consisted of predominantly rural population with mixed socio-economics status.

\subsubsection{Data collection and evaluation}

The data collection took place by a pre-test, classroom observations of the
group work, worksheets and tests. The pre-test had 28 short questions. The
first part ($7$ questions) was related to a general data about a student:
gender, age, secondary school, final exam, residence stratum and motivation
for science subjects. The second part ($19$ questions) was related to liquid
crystals, their existence, properties and microscopic structure. The
pre-test was applied at the beginning of lecture related to liquid crystals.
Those students who did not attend the lecture filled in the pre-test before
the beginning of the compulsory lab work in chemistry.

The worksheet for the lab work in chemistry includes a procedure to
synthesise liquid crystal MBBA, a reaction scheme, observations and
conclusions regarding the synthesis and questions from chemistry related to
liquid crystals and the lab work.

The worksheet for the lab work in physics presents properties of polarizing
foils and optically anisotropic materials and experiments with the liquid
crystal MBBA.

Test $1$ includes $17$ short questions related to the knowledge obtained during
the lecture and lab work. Test $1$ was held immediately after the end of the
physics lab (in May 2011).

Test 2 was a part of an exam held $4$ weeks later (June 2011). It has $17$
questions that, again, cover the contents of the lecture and lab work.
Questions on test $2$ were similar to questions given on the pre-test and test
$1$.

The study provided an extensive set of data but in this paper we will focus
only on students' comprehension of new concepts.

\subsection{Results and discussion}

Results of the pre-test show that $94.4\ \%$ of students have already heard of
liquid crystals. The percentage is so high, because we were testing students
informally obtained knowledge about liquid crystals as a part of another
study held at the beginning of the academic year. One student said:
\textquotedblleft When we got the questionnaire at the beginning of the
academic year I was ashamed because I did not know anything about liquid
crystals. When I came home I asked my father and checked on the web what
they are. These experiments definitely bring them closer to
me.\textquotedblright\ Such an interest is a rare exception, however, most
of the students remembered the term \textquotedblleft liquid
crystals\textquotedblright , which was a central point of the questionnaire
that they filled in at the beginning of the academic year 2010/11.

Since lectures are not compulsory only $37$ students attended the lecture. 150
students attended compulsory laboratories. They worked on the synthesis a
week after the lectures and another week later on experiments with liquid
crystals at the physics lab. Students worked in groups of $3$ or $4$ in the
chemistry lab and in pairs in the physics lab. However, the whole data
(tests and worksheets) was collected only for 90 students, therefore we
present only their achievements.

All the groups made the synthesis successfully according to the procedure
written in the worksheets. $40$ syntheses out of $40$ were successfully carried
out which was confirmed by measuring the melting and clearing temperature of
the synthesized liquid crystal MBBA. On average $63.3\ \%$ of worksheets were
correctly filled in ($SD=13.0\ \%$).

All the experiments described in section 2.4.3 were successfully carried out
in the physics lab. The only difference was that students did not prepare
the wedge cell by themselves. Due to the lack of time cells were prepared in
advance. On average 84 \% of worksheets included correct answers to
questions and observations ($SD=9\ \%$).

On the pre-test students on average achieved $24.0\ \%$ of all points. Their
achievements show that their prior knowledge about liquid crystals was
limited, as expected. On test 1 that was held immediately after the physics
lab, students on average achieved 68.1 \% (see table \ref{table:tableI}).
Test $2$ was a part of a regular exam in Physics. On test $2$ students on
average achieved only $63.5\ \%$ points. The reduced performance on test 2 can
be explained by the research on memory and retention,
which suggests that many standard educational practices, such as exams and a
great emphasis on the final exam, which encourages studying by cramming, are
likely to lead to the enhanced short-term performance at the expense of a
poor long-term retention \cite{Bjork_1994}. 
\begin{table}[h]
\caption{ Students achievements on tests }
\label{table:tableI}{\scriptsize 
\begin{tabular}{p{2 cm}p{3 cm}p{0.5 cm}p{2 cm}p{2cm}}
\br Test & Average number of achieved points (max) & ~ & $\mathit{SD}$ & 
Percentage of achieved points \\ 
\mr Pre-test & 6.0 (25) & ~ & 2.9 & 24.0 \\ 
Test 1 & 14.0 (20.5) & ~ & 2.4 & 68.1 \\ 
Test 2 & 14.0 (22) & ~ & 2.6 & 63.5 \\ 
\br &  &  &  & 
\end{tabular}
}
\end{table}

\begin{figure}[h]
\centering \includegraphics[scale=0.6]{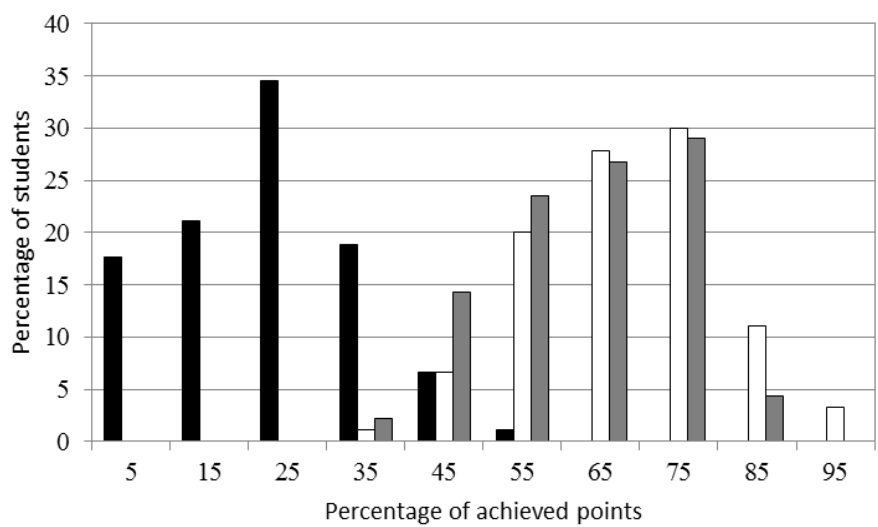} 
\caption{Distribution of the percentage of students vs. percentage of
achieved points on tests. Black: pre-test; white: test 1; grey: test 2. }
\label{fig:Fig_7}
\end{figure}

Figure \ref{fig:Fig_7} shows the distribution of students vs. the achieved
points on the pre-test, test $1$ and test $2$. The percentage of students who
achieved higher scores on tests $1$ and $2$ with respect to the pre-test is
evident. The expected level of knowledge about liquid crystals is the
highest immediately after the activities. From figure \ref{fig:Fig_7} it is
seen that most students achieved less than $50\ \%$ of points on the pre-test.
On test $1$ most of the students achieved over $50\ \%$. It is clearly seen that
the percentage of students achieving higher percentage of points on test $2$
is lower than on test $1$, when the impressions of the lab work were still
fresh in mind. 

A comparison of the percentage of correct answers on selected questions from
the pre-test, test 1 and test 2 (table \ref{table:tableII}) shows further
details of students comprehension. Questions on tests were not always
exactly the same, but they covered the same contents. The results show that
the percentage of students who answered correctly is higher on tests 1 and 2
in comparison to the pre-test, which shows the efficiency of the teaching
module.

On test 1, $73\ \%$ of students correctly answered that liquid crystals are a
state of matter. The percentage of students who answer correctly this
question on test 2 is smaller. It seems that the deeply rooted misconception
about the existence of only three states of matter prevailed after some
period of time over new concepts met only during the teaching about liquid
crystals.

As much as $93\ \%$ of students stated at least one product with liquid
crystals on test 2. The percentage of students who underlined correctly all
three properties of liquid crystals which are important for their
applications, i.e. colours, birefringence and electric properties, is the
highest on test 2.

On both tests approximately $60\ \%$ of students agreed with the statement that
liquid crystals are also in living organisms and $80\ \%$ knew that double
refraction can be observed in anisotropic materials.

Liquid crystals are liquid in the liquid crystalline state was a statement
with which $76\ \%$ of students agreed on both tests.

On test 2, $88\ \%$ of students knew that an electric field has an effect on
the liquid crystal molecules.

The biggest jump in knowledge was detected in the question related to the
propagation of light in a birefringent material. On test 1, $98\ \%$ of
students correctly sketched that the laser beam splits into two beams in
birefringent materials. However, the percentage of students who assimilate
this was lower, only $58\ \%$. The concept is very difficult and students did
not have any preliminary knowledge about it. However, experiments were
straightforward in showing that a single light beam splits into two, which
is consistent with the phrase \textquotedblleft double refraction\textquotedblright.

Finally, $100\ \%$ of students sketched the distribution of molecules in the
liquid crystalline state correctly on test 1 while the percentage was
reduced to $73\ \%$ on test 2. The result is certainly again influenced by the
lab work, where the reasons for the anisotropic properties based on the
microscopic structure were discussed in detail.

The results of tests 1 and 2 show that fresh impressions from the lab faded
away by test 2. This is expected for rather disinterested students that,
unfortunately, are rather common in this specific study program.
Nevertheless, the results offer an interesting starting point for more
extensive research on retention with respect to interest in different
physics phenomena. 
\begin{table}[h]
\caption{The percentage of correct answers per question on the pre-test and
tests}
\label{table:tableII}{\scriptsize 
\begin{tabular}{p{6cm}p{0.5cm}p{1.50cm}p{1.50cm}p{1.5cm}}
\br \multicolumn{2}{l}{} & \multicolumn{3}{l}{Percentage of student who answered
correctly} \\ 
Question & ~ & Pre-test & Test 1 & Test 2 \\ 
\mr Liquid crystals are a state of mater. & ~ & 16 & 73 & 61 \\ 
& ~ & ~ & ~ &  \\ 
Write down a product with liquid crystals. & ~ & 38 & 86 & 93 \\ 
& ~ & ~ & ~ &  \\ 
Which properties are most important for the application of liquid crystals?
You can choose more than one property. & ~ & 34 & 76 & 87 \\ 
& ~ & ~ & ~ &  \\ 
Substances with liquid crystalline properties appear in living organisms. ~
&  & 18 & 61 & 60 \\ 
& ~ & ~ & ~ &  \\ 
Double refraction can be observed in the anisotropic materials. & ~ & 18 & 81
& 82 \\ 
& ~ & ~ & ~ &  \\ 
Liquid crystals are solid in the liquid crystalline state. & ~ & 43 & 76 & 76
\\ 
& ~ & ~ & ~ &  \\ 
Electric field can influence the orientation of the liquid crystal molecules.
& ~ & 49 & 67 & 88 \\ 
& ~ & ~ & ~ &  \\ 
Draw the light propagation from air to water and from air to the
birefringent material. & ~ & 2 & 98 & 58 \\ 
& ~ & ~ & ~ &  \\ 
Draw the distribution of molecules in liquid crystals. & ~ & 39 & 100 & 73
\\ 
\br &  &  &  & 
\end{tabular}
}
\end{table}

The evaluation of the teaching module shows that it efficiently increases
the knowledge about liquid crystals. By comparing the results of the
pre-test, test 1 and test 2 one can conclude that the teaching module was
appropriately designed and it allows for the development of concepts related
to liquid crystals (table \ref{table:tableII}). Students were actively
involved in the learning process. Their engagement was the lowest during the
lecture. At the chemistry lab work students worked in groups of 3 or 4 in a
lab with only 2 fume hoods. Due to the waiting for hoods during the
synthesis of MBBA not all the members of a group were fully focused on the
synthesis. There were also 20 minutes of \textquotedblleft free\textquotedblright time due to the time needed
for the reaction. So, one should consider the possibilities of better
organization of the chemistry lab work in order to keep the students focused
on the work. At physics lab students worked in pairs and were more active
throughout. It has to be pointed out again that mostly female students were
included in the study, the fact that might affect the generality of the
conclusions. However, the study confirms that students in general
assimilated the most important concepts related to liquid crystals and their
application. The results of the tests confirm that, after the teaching
module, students were not only aware of liquid crystals but they also
learned their relevant properties.

\section{Discussion and conclusions}

Liquid crystals are materials that flow like liquids and have physical
properties of solid crystals. They are quite common both in nature and
technology where they are used in laptops, mobile phones, mp4-players etc.
At the same time liquid crystals are an important topic in current
scientific research. Liquid crystals are therefore a topic which fulfils two
major conditions for being relevant and motivating for students. Because of
that we have designed a teaching module which has three parts: the lecture,
chemistry lab and physics lab. The aim of the module is to give students a
general knowledge about liquid crystals, their properties and the principles
of how LCDs work.

The lecture gives the basic knowledge about liquid crystals and their
properties. With the lab work students strengthen and expand their
knowledge. They synthesize liquid crystal MBBA, they use their own product,
i.e. MBBA, in experiments where they observe and discuss an additional phase
transition, the properties of optically anisotropic materials in general,
double refraction and colours observed in liquid crystalline cells.

The module was tested by 90 first-year, mostly female, students in the study
program for teachers for the lower grades of elementary school in the
academic year 2010/11. These pre-service teachers have no preference to
natural sciences. The knowledge obtained by the teaching module was
confirmed by results of the tests. Students on average achieved $68.1\ \%$ on
the test immediately after the activities and $63.5\ \%$ on the test which was
a part of an exam a month later. The achievements show a significant
increase in the knowledge about liquid crystals, since at the pre-test
students achieved on average only $24.0\ \%$.

Can the results of the study for this specific group be generalized to a
general population of the first year university students? The group
consisted of future primary school teachers and it is not obvious that one
can make general assumptions based on the results obtained for this group.

In order to find out if the results obtained by this group can lead to
general conclusions we performed additional studies prior to the study reported in this paper. We used a liquid
crystal questionnaire (LCQ) to test the informally obtained
pre-knowledge about liquid crystals of the first year students at various
study programs at the University of Ljubljana \cite{eurasia}. Within the LCQ we assessed
other circumstances like the intellectual level of the first year students
at different representative study programs in order to estimate the
equivalence of the tested and a general group. The results of the study
performed on a more general and wider sample of $1121$ students shows that a
general student pre-knowledge about liquid crystals is practically
negligible (table \ref{table:tableIV}), although male students showed
statistically significant better achievements on LCQ than female students.
This fact is in agreement with the findings of Haeussler and Hoffmann \cite%
{Haeussler_2002}, which argue that male students are more interested in
technology and application of science than female students. 
\begin{table}[h]
\caption{Comparison between the achievements on LCQ of pre-service primary
school teachers from Faculty of Education (pilot study group), students from
Faculty of Education and students from University of Ljubljana}
\label{table:tableIV}{\scriptsize 
\begin{tabular}{p{3cm}p{1cm}p{1.3cm}p{2.1cm}p{1.3cm}p{2.1cm}p{1.4cm}p{2.1cm}}
\br &  & \multicolumn{2}{l}{Pilot study group} & \multicolumn{2}{l}{Faculty
of Education} & \multicolumn{2}{l}{University of Ljubljana} \\ 
&  & \multicolumn{2}{l}{($\mathit{n}=82$)} & \multicolumn{2}{l}{($\mathit{n%
}=278$)} & \multicolumn{2}{l}{($\mathit{n}=1121$)} \\ 
\mr Achievements on the LCQ (No. of points out of 8) & ~ & 1.3 & ($\mathit{SD%
}=1.4$) & 1.4 & ($\mathit{SD}=1.4$) & 1.9 & ($\mathit{SD}=1.5$) \\ 
~ & ~ & ~ & ~ & ~ & ~ & ~ &  \\ 
Achievements on the final exam (average in Slovenia: 19.5 points out of 34)
& ~ & 19.0 & ($\mathit{SD}=3.9$) & 19.5 & ($\mathit{SD}=4.7$) & 22.0 & ($%
\mathit{SD}=5.1$) \\ 
~ & ~ & ~ & ~ & ~ & ~ & ~ &  \\ 
Gender and achievements on LCQ & Male & 6 \% & 1.8 ($\mathit{SD}=1.3$) & 10 \% & 2.4 ($\mathit{SD}= 1.3$) & 32 \% & 2.7 ($\mathit{SD}=1.5$) \\ 
& Female & 94 \% & 1.25 ($\mathit{SD}=1.4$) & 90 \% & 1.3 ($\mathit{SD}=1.4$) & 68 \% & 1.5 ($\mathit{SD}=1.4$) \\ 
& ~ & ~ & ~ &  &  &  &  \\ 
& $\mathit{t}$-test & $\mathit{t}=0.840$ & $\mathit{p}=0.404$ & $\mathit{%
t}=3.691$ & $\mathit{p}= 0.000$ & $\mathit{t}=13.371$ & $\mathit{p}=0.000$ \\ 
\br &  &  &  &  &  &  & 
\end{tabular}
}
\end{table}

The difference between the pre-service students and students from faculties
where the main study field is science and technology is in motivation for
natural sciences. The students in the study fields connected to natural
sciences and technology definitely have higher interest in natural sciences
and therefore it is somehow expected that their achievements on LCQ will be
higher. The assumption was confirmed. There are statistically important
differences in achievements on LCQ between the students of natural sciences
and technology and those who study social sciences and humanistic (see table %
\ref{table:tableV}).

\begin{table}[h]
\caption{Achievements on the LCQ regarding the field of study and the
natural sciences fields}
\label{table:tableV}{\scriptsize 
\begin{tabular}{p{2.2 cm}p{0.1 cm}p{2.0 cm}p{0.1cm}p{3.5 cm}p{0.1cm}p{1cm}p{1 cm}p{1cm}}
\br Field of study ($\mathit{n}= 1121$) & ~ & percentage of students & ~ & 
number of achieved points out of 8 & ~ & $\mathit{SD}$ & $\mathit{t}$ & $%
\mathit{p}$ \\ 
\mr natural sciences or technology & ~ & 41.7 & ~ & 2.5 & ~ & 1.6 & ~ & ~ \\ 
~ & ~ & ~ & ~ & ~ & ~ & ~ & 12.712 & 0.000 \\ 
social sciences or humanistic & ~ & 58.3 & ~ & 1.4 & ~ & 1.6 & ~ & ~ \\ 
\br &  &  &  &  &  &  &  & 
\end{tabular}
}
\end{table}

To conclude, the results from the testing of informally gained knowledge at
the end of the secondary school of the pre-service teachers and the students
from different faculties on average do not differ: in both samples the lack
of knowledge about liquid crystals was detected. Based on the research of
prior knowledge it can be said that students that are interested in natural
sciences and technology would assimilate at least as much knowledge about
liquid crystals from the teaching module as the pre-service teachers did.
Since the results of prior knowledge show statistically significant
differences between the knowledge of male and female students one can even
dare to conclude that the male students would achieve better results on
testing of the module as female students. One can therefore safely conclude
that a general audience would achieve at least as good results as the group
of students involved in this study. It must be stressed that we have
designed the module in which students acquire new knowledge relevant to
liquid crystals and their applications, as confirmed by the implementation
and the evaluation of the module. Module can also be used as a teaching
module at more specialized physics courses. Of course there are still opened
issues that we intend to explore. Since we know that the module is
appropriate for students that are not motivated in science we will work on
the adaptation of module for students motivated in natural sciences. The
module presented in this paper can be readily used in the introductory
physics courses, and with appropriate modifications it can also be used at
lower levels of education.

Evaluation of the model raised several questions that need to be addressed
in the future: How do practical experiences influence the knowledge about
liquid crystals?, How does a new learning environment influence the
knowledge?, How do the chosen teaching methods influence the study process?,
etc. However, the whole study and its evaluation show that it is worth an
effort to develop new modules on topics related to the current scientific
research and everyday technology.

\ack{We are grateful to the students who were part of the research, especially to the pre-service primary school teachers from Faculty of Education (University of Ljubljana).

The presented work was partialy  funded by the Slovenian Research Agency (ARRS) within the project J5-4002. }

\end{document}